\documentclass{article}

\usepackage{graphicx}

\begin{document}

\title{Effects of Contact Parameters on Transmission through a Benzene Ring}

\date{August 29, 2019}

\author{Kenneth W. Sulston\thanks{email: sulston@upei.ca} 
\thanks{School of Mathematical and Computational Sciences, University of Prince Edward Island, Charlottetown, PE, C1A 4P3, Canada}}

\maketitle

\section{Abstract}

Transmission through a benzene ring, connected by sulfur contact atoms to gold leads, is calculated by a tight-binding
model by means of the renormalization method. 
Attention is focused on the parameters associated with the contact atoms, namely their site energy
and their bond energies with the ring and the leads.
These parameters are found to have significant effects on the transmission probability function.

\section{Introduction}

\begin{figure}[htbp]
\includegraphics[width=12cm]{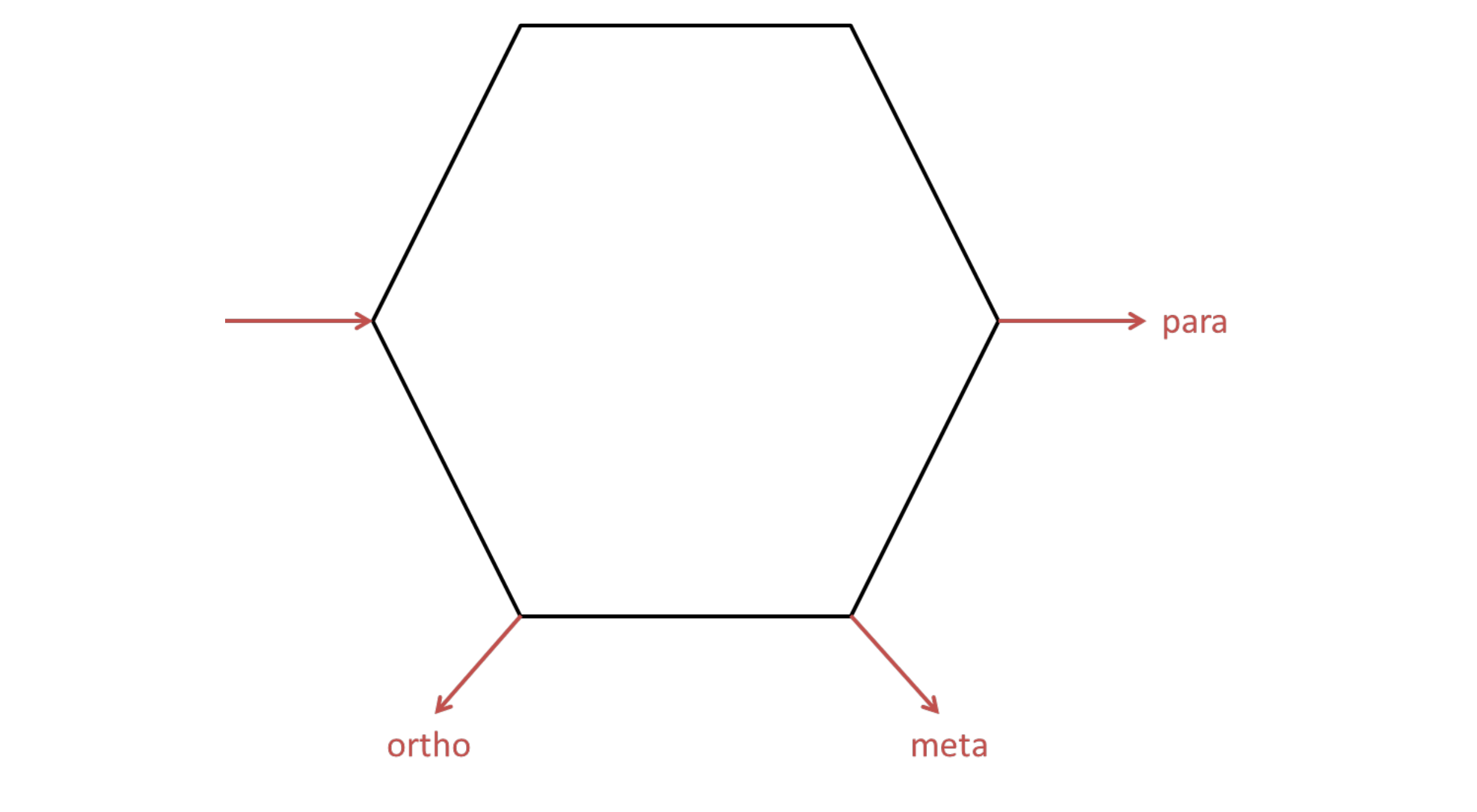}
\caption{Benzene molecule showing para, meta and ortho connections to leads.}
\label{fig1}
\end{figure}

Benzene is an important molecule for many reasons, among which is its potential applicability to 
electron transport in single-molecule devices, either in its own right or as a tool to understanding more
complicated systems, such as those involving aromatic hydrocarbons \cite{ref1}, graphene \cite{ref2}, 
carbon nanotubes, etc.
Thus, an understanding of its electron-transmission properties is a basis for exploring much of molecular electronics.
One relevant study was that by Kucharczyk and Davison \cite{ref3}, who used a renormalization method \cite{ref4, ref5} 
(which is also used in the current paper)
to calculate the local density of states over a benzene molecule attached to polymeric chains, 
with attention paid to the differences among the para, meta and ortho configurations. (see Figure \ref{fig1}).
One analytical approach was by Hansen {\it et al} \cite{ref6},
who used the L\"{o}wdin partitioning technique to derive the Green's function for a benzene ring.
The coupling of the molecule to metallic leads was then described using first-order
 perturbation theory, whence the transmission probability $T(E)$, as a function of energy $E$, could be calculated.
Another analytical approach was by Dias and Peres \cite{ref7}, who used the Green's function method, within
the tight-binding approximation, to derive an expression for the transmission function $T(E)$
through a para-benzene ring.

Our studies of benzene systems utilise the afore-mentioned renormalization method \cite{ref4, ref5}
to calculate Green's functions within the tight-binding approximation. 
The transmission function is then evaluated via the Lippmann-Schwinger equation.
The foundational paper of this group \cite{ref8} details the methodology, and considers transmission through
 a single benzene ring in para, meta and ortho configurations, as well as these rings assembled into series and
 parallel circuits. 
 The method was extended \cite{ref9} to include overlap effects, which were seen to be considerable, and in
 particular, had the effect of breaking the symmetry in the $T(E)$ curves.
 Another study \cite{ref10} found analytic formulas for the transmission functions $T(E)$ of single benzene 
 molecules, and was shown to be in agreement with other work \cite{ref6, ref7}.
 
In the current work, we extend and fine-tune our previous work on single benzene rings, by modelling
the ring as being attached to sulfur contacts, which in turn connect to gold leads. 
We concentrate on how the transmission $T(E)$ depends on the contact parameters, and the
effect on transmission of varying these parameters.
In this way, we can gauge, to some extent, the sensitivity of the transmission through benzene upon the
specifics of the ring's connection to its leads.

\section{Model}

\begin{figure}[htbp]
\includegraphics[width=12cm]{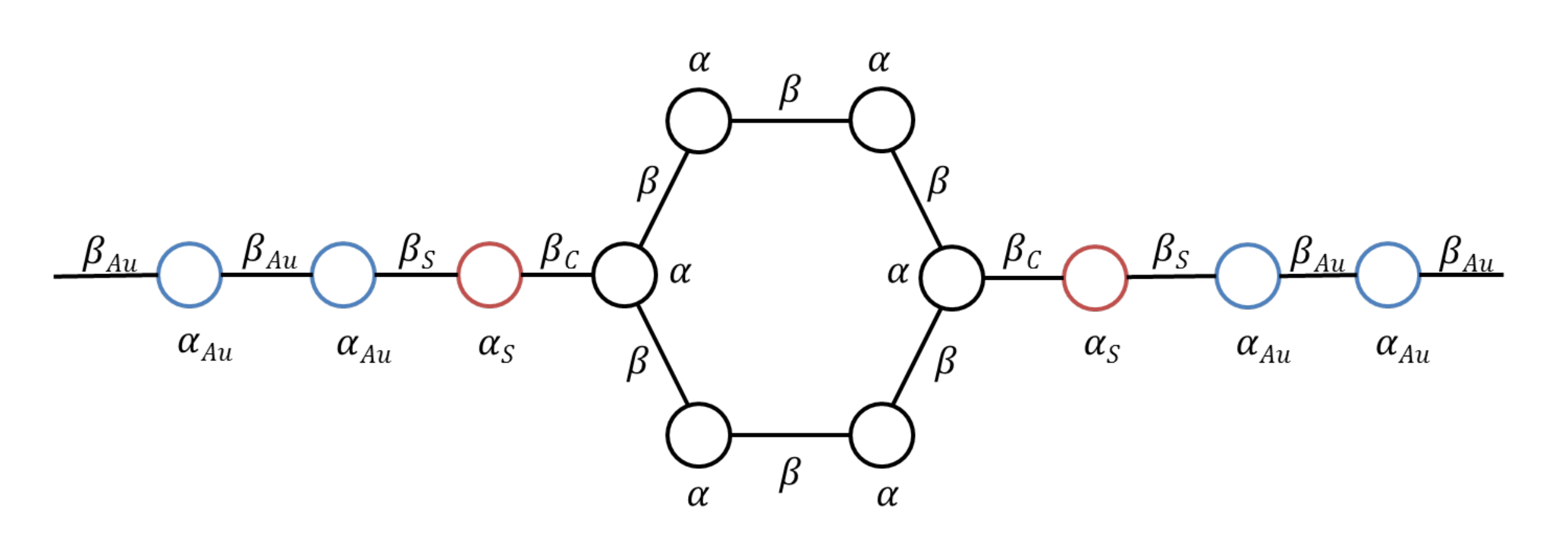}
\caption{Benzene molecule (showing C atoms only), with S contacts connected to Au chain leads.}
\label{fig2}
\end{figure}

The model under consideration is shown in Figure \ref{fig2}.
The benzene molecule considers carbon atoms only within the tight-binding approximation, with site energy
$\alpha$ and bond energy $\beta$.
Two of the carbon atoms, depending on the specific configuration, are attached to sulfur atoms by bonds $\beta_C$.
The sulfur atoms have site energy $\alpha_S$, and are in turn attached to gold atoms by bonds $\beta_S$.
One-dimensional chains of gold atoms serve as the leads, with the gold atoms having site energy $\alpha_{Au}$
and bond energy $\beta_{Au}$.
As we wish to investigate the effects of the contacts, the parameters $\alpha_S$, $\beta_C$ and $\beta_S$
are of primary interest, so we look at how variation in their values affects the $T(E)$ curves.
For ``standard'' parameter values, we use \cite{ref11, ref12} $\alpha=-6.553$ eV, $\beta=-2.734$ eV, $\alpha_{Au}=-8.43$ eV,
$\beta_{Au}=-1.325$ eV, $\alpha_S=-6.553$ eV, $\beta_C=-2.1872$ eV and $\beta_S=-1.325$ eV.

\begin{figure}[htbp]
\includegraphics[width=12cm]{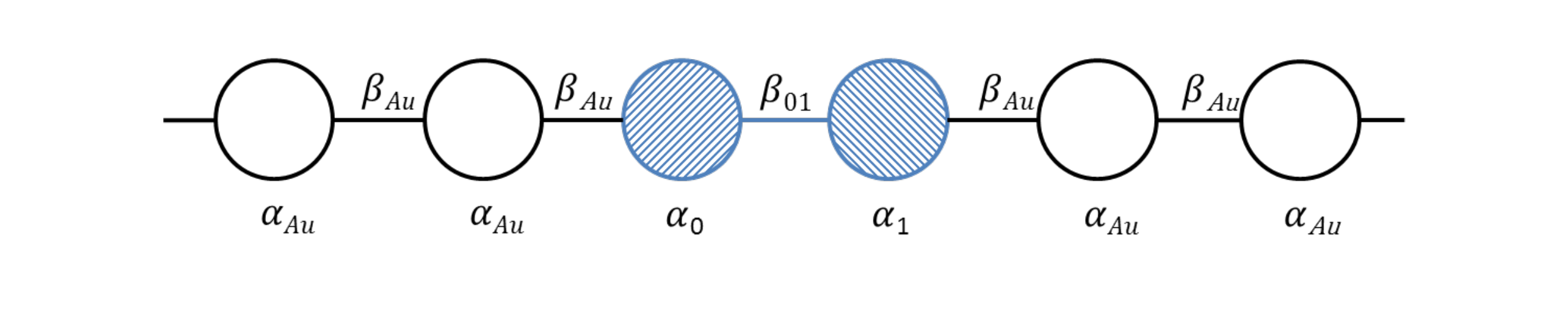}
\caption{Semi-infinite Au chain leads attached to a dimer, with rescaled site energies $\alpha_0$ and $\alpha_1$ and rescaled
bond energy $\beta_{01}$.}
\label{fig3}
\end{figure}

Our starting point is a brief summary of the relevant results from \cite{ref8} for transmission
through a single benzene molecule, without specialized contacts;
for complete details, we refer the reader to \cite{ref8}.
The basic method therein was to use the Lippmann-Schwinger equation to derive the 
transmission probability for an electron through a one-dimensional tight-binding chain,
containing a double impurity, as shown in Figure \ref{fig3}.
Subsequently, this allows benzene (and indeed, a wide variety of systems) to be studied,
by using the renormalization technique to reduce the molecule to a dimer, with rescaled
energy-dependent parameters, which then plays the role of the double impurity.
The main mathematical result is that the transmission-energy probability function has the
form
\begin{equation}
T(X(E))= {{(1+2\gamma)^2 (4-X^2)} \over {(1-2Q)^2 (4-X^2) + 4(P-QX)^2}}  ,
\label{eq1}
\end{equation}
where
\begin{equation}
P = z_0+z_1  ~,~ Q = z_0 z_1 - \gamma -\gamma^2 ,
\label{eq2}
\end{equation}
with
\begin{equation}
z_{0,1} = ({\alpha}_{0,1} - \alpha_{Au})/ 2\beta_{Au} ~,~ \gamma = ({\beta}_{01} -\beta_{Au})/ 2\beta_{Au} ,
\label{eq3}
\end{equation}
and the reduced dimensionless energy is
\begin{equation}
X= (E - \alpha_{Au})/\beta_{Au}.
\label{eq4}
\end{equation}
In the above, $\alpha_{Au}$ and $\beta_{Au}$ are the site and bond energies, respectively, for
the gold atoms in the leads, while $\alpha_0$, $\alpha_1$ and
$\beta_{01}$ are the rescaled parameters for the dimer.
In general situations, the dimer may be asymmetric resulting in $\alpha_0 \ne \alpha_1$, but
for the cases  considered here, the dimers are all symmetric, so that
$\alpha_0 = \alpha_1$, and hence $z_0=z_1$ in (\ref{eq3}).
The strength and flexibility of the renormalization method lies in its ability to reduce quite complicated molecules, atom by atom,
to dimers with rescaled parameters, whose transmission function is then given by (\ref{eq1}).

In comparing the current model shown in Figure \ref{fig2} to the general situation of Figure \ref{fig3}, we see that we must
use renormalization to reduce the benzene molecule plus the contacts plus the adjoining gold atoms to a dimer, 
in order to utilize (\ref{eq1}).
We refer again to \cite{ref8}, where the (symmetric) dimers for the three benzene configurations were calculated. 
On letting
\begin{equation}
Y = (E-\alpha)/\beta ,
\label{eq5}
\end{equation}
the results are as follows.
For para-benzene,
\begin{equation}
\bar{\alpha}_p  = \alpha + \bar{\beta}_p Y ,
\label{eq6}
\end{equation}
\begin{equation}
\bar{\beta}_p = 2 \beta (Y^2-1)^{-1} .
\label{eq7}
\end{equation}
For meta-benzene,
\begin{equation}
\bar{\alpha}_m  = \alpha + \beta Y^{-1} + \bar{\beta}_m ,
\label{eq8}
\end{equation}
\begin{equation}
\bar{\beta}_m =  \beta Y^{-1} (Y^2-1) (Y^2-2)^{-1}.
\label{eq9}
\end{equation}
For ortho-benzene,
\begin{equation}
\bar{\alpha}_o  = \alpha + \beta (Y^2-2) (Y^2-Y-1)^{-1} - \bar{\beta}_o ,
\label{eq10}
\end{equation}
\begin{equation}
\bar{\beta}_o =  \beta (Y^2-1) (Y^2-2) [(Y^2-1)^2-Y^2]^{-1}.
\label{eq11}
\end{equation}

With the benzene ring now replaced by a dimer in  the chain of Figure \ref{fig2}, the process is now to use the renormalization
method to reduce that dimer, its sulfur contacts, and the immediately-adjacent gold atoms (because they are connected to sulfur atoms
by $\beta_S$ not $\beta_{Au}$) to a new dimer, embedded within a chain of gold atoms, of site energy $\alpha_{Au}$ and bond
energy $\beta_{Au}$, matching the set-up of Figure \ref{fig3}.
Then equation (\ref{eq1}) is applicable to calculate $T(E)$.
The renormalization method acts to reduce this part of the chain to a dimer, by decimating an atom from the chain while rescaling
the site and bond parameters on the adjacent atoms. 
It proceeds atom-by-atom, first on each sulfur contact atom, and then on each gold atom at the end of a lead.
Specifically, the renormalization equations to remove the atom at site ``$n$'' are \cite{ref8}
\begin{equation}
\tilde{\alpha}_{n-1} =\alpha_{n-1}+ { {\beta}_{n-1,n}^2 \over {E-\alpha_n}},
\label{eq12}
\end{equation}
\begin{equation}
\tilde{\alpha}_{n+1} =\alpha_{n+1}+ { {\beta}_{n,n+1}^2 \over {E-\alpha_n}},
\label{eq13}
\end{equation}
\begin{equation}
\tilde{\beta}_{n-1,n+1}  = {{\beta_{n-1,n} \beta_{n,n+1}} \over {E-\alpha_n}}.
\label{eq14}
\end{equation}
After repetitive application of these equations to the 4 atoms to be removed, the resulting dimer has rescaled parameters
\begin{equation}
\alpha_0 = \alpha_1 = \alpha_{Au}+ { \beta_S^2 [(E-\bar{\alpha})^2-\bar{\beta}^2]
 \{ (E-\alpha_S) [(E-\bar{\alpha})^2-\bar{\beta}^2] - \beta_C^2(E-\bar{\alpha}) \} \over
  {\{(E-\alpha_S) [(E-\bar{\alpha})^2-\bar{\beta}^2] - \beta_C^2(E-\bar{\alpha})\}^2-\beta_C^4 \bar{\beta}^2}},  
\label{eq15}
\end{equation}
\begin{equation}
\beta_{01}  = {{\beta_S^2 \beta_C^2 \bar{\beta}^2 [(E-\bar{\alpha})^2-\bar{\beta}^2] } \over 
{\{(E-\alpha_S) [(E-\bar{\alpha})^2-\bar{\beta}^2] - \beta_C^2(E-\bar{\alpha})\}^2-\beta_C^4 \bar{\beta}^2}}. 
\label{eq16}
\end{equation}
In (\ref{eq15}) and (\ref{eq16}), $\bar{\alpha}$ and $\bar{\beta}$ can be subscripted by $p$, $m$ or $o$, as desired,
and the corresponding results of (\ref{eq6})-(\ref{eq11}) utilised, 
to produce the para, meta or ortho configuration, respectively.

\section{Results and Discussion}

In this section, we look at a selection of $T(E)$ curves, calculated via (\ref{eq1}), based on the ``standard'' parameters given at the
beginnning of section 3. 
Specifically, we take as fixed the parameters $\alpha=-6.553$ and $\beta=-2.734$ for the C atoms in the benzene ring, and 
$\alpha_{Au}=-8.43$ and $\beta_{Au}=-1.325$ for the Au atoms in the leads. 
As it is the effect of the S contacts that we wish to examine, the relevant parameters $\alpha_S=-6.553$, $\beta_C=-2.1872$
and $\beta_S=-1.325$ are allowed to vary, in turn, from their standard values, for each of the three benzene configurations.
In each of the figures presented in the following subsections, the green solid curve of (a) uses the standard parameters, while
the other curves (b)-(d) are for varying values of the parameter being considered.

\subsection{Para-benzene}

\begin{figure}[htbp]
\includegraphics[width=12cm]{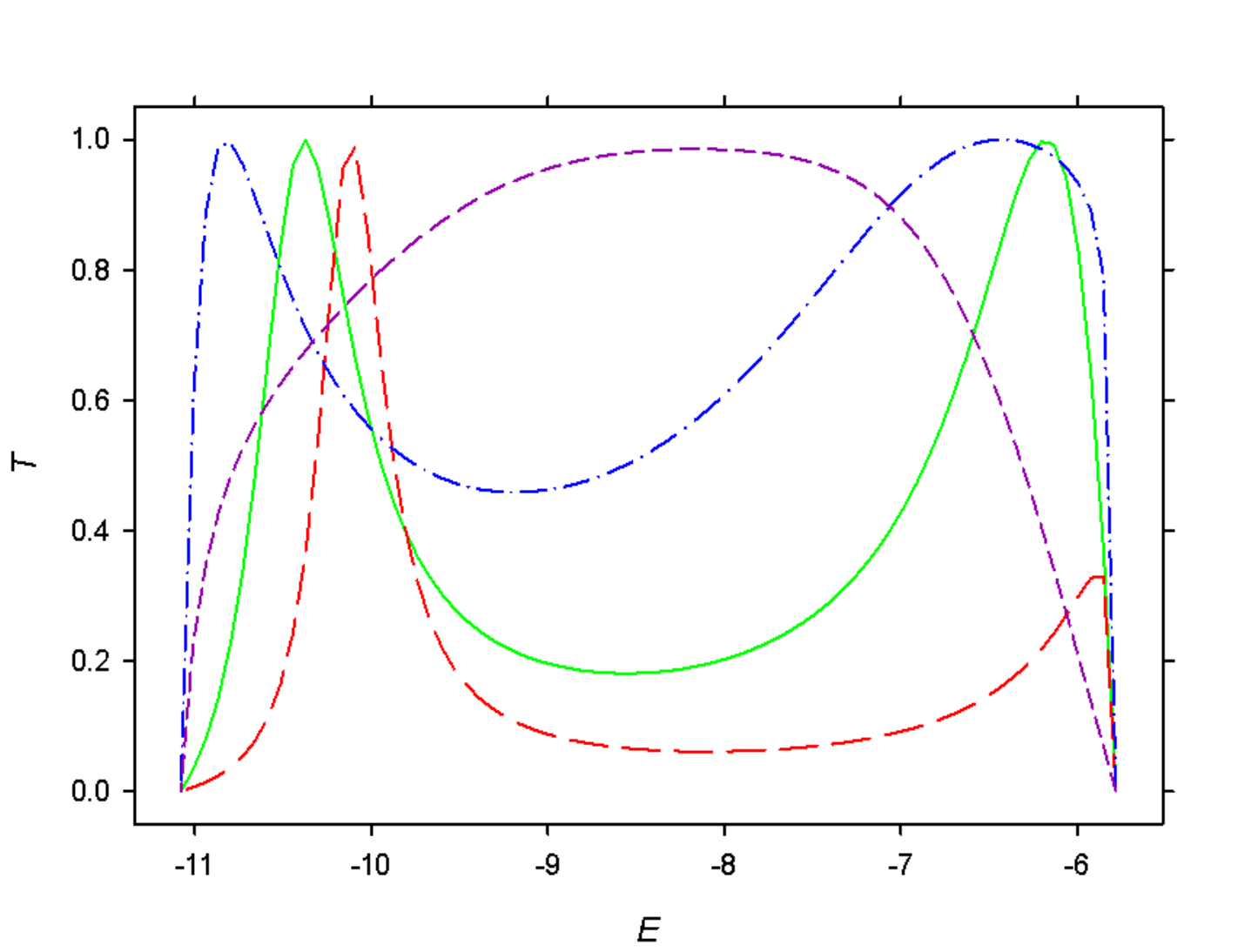}
\caption{Transmission $T$ versus energy $E$ for para-benzene, with $\alpha_S=$ (a) $-6.553$ (green solid curve), 
(b) $-5.5$ (red long-dashed), (c) $-7.5$ (blue dash-dotted), (d) $-8.5$ (purple short-dashed). }
\label{fig4}
\end{figure}

We start by looking at para-benzene, and the dependence of its $T(E)$ curve on the contact parameters. 
Figure \ref{fig4}(a) displays the $T(E)$ curve for the standard value of $\alpha_S=-6.553$, in which case the curve
is almost symmetric, with a local minimum of $T \approx 0.2$ near the band center, flanked by a pair of resonances 
(for which $T=1$) close to the band edges. 
At the band edges themselves, and only there, the transmission $T=0$ exactly (for all parameter values considered in this subsection).
When $\alpha_S$ is raised, as exemplified in Figure \ref{fig4}(b) with $\alpha_S=-5.5$, the local minimum persists and indeed
goes lower, but the upper resonance disappears with the peak becoming progressively smaller as $\alpha_S$ increases. 
Meanwhile the lower resonance survives, albeit at slightly higher $E$, but becomes increasingly narrow and eventually approaches
a $\delta$-function with further increases in $\alpha_S$.
So in this case, transmission at lower energies is clearly preferred.
On the other hand, lowering $\alpha_S$ to $-7.5$ (Figure \ref{fig4}(c)) maintains both resonances and broadens both, especially 
the upper one, and correspondingly raises the local minimum, thus enhancing transmission significantly across most of the band.
Further lowering of $\alpha_S$ to $-8.5$ (Figure \ref{fig4}(d)), which is very close to $\alpha_{Au}$, produces a very different 
curve, with the local minimum completely gone and replaced by a single resonance, with broad width.
As $\alpha_S$ is lowered further still (not shown), the single resonance splits back into two resonances separated by a local
minimum, very similar to the curves shown in (a) and (b).
Thus there is a quite noticeable effect of having the site bond $\alpha_S$ of the contact atom being very different from that
of the chain atoms.

\begin{figure}[htbp]
\includegraphics[width=12cm]{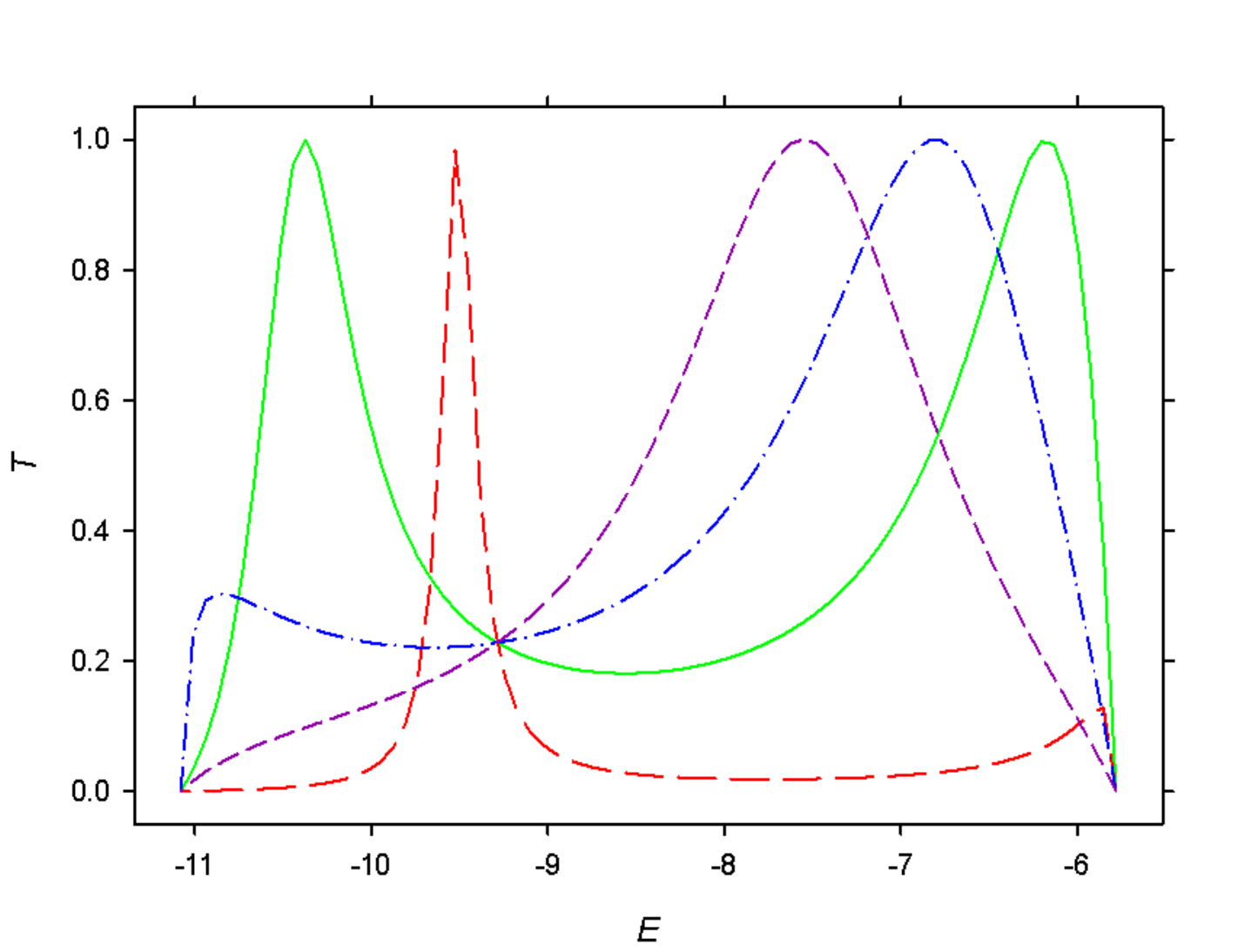}
\caption{Transmission $T$ versus energy $E$ for para-benzene, with $\beta_C=$ (a) $-2.1872$ (green solid curve), 
(b) $-1.0$ (red long-dashed), (c) $-3.0$ (blue dash-dotted), (d) $-4.0$ (purple short-dashed). }
\label{fig5}
\end{figure}

Turning next to the effect of $\beta_C$, we start again from the $T(E)$ curve for the standard parameters, including
$\beta_C=-2.1872$, shown earlier in Figure \ref{fig4}(a) and reproduced in Figure \ref{fig5}(a).
Again this curve is dominated by the two resonances separated by a local minimum.
Lowering $|\beta_C|$ towards 0 (i.e., increasing $\beta_C$), as shown in Figure \ref{fig5}(b), has the overall effect 
of inhibiting transmission.
In particular, the upper resonance is drastically reduced, becoming merely a small peak near the upper band edge.
The exception is the lower resonance, which persists at a somewhat higher energy and with significant narrowing.
Further reductions in $|\beta_C|$ (not shown) continue these effects, reducing transmission to virtually 0 at all energies
except near the lower resonance, which further narrows into a near $\delta$-function at energy at around $-9.3$.
Increasing $|\beta_C|$ has a very different (and opposite) effect however. 
Taking $\beta_C=-3.0$, as shown in Figure \ref{fig5}(c), greatly diminishes the {\it  lower} resonance while maintaining
and broadening the upper one, albeit at a slightly lower energy.
This effect is enhanced by further lowering of $\beta_C$ to $-4.0$ (see Figure \ref{fig5}(d)) with the one-time lower resonance
no longer distinguishable but the upper resonance further broadened.
This results in generally stronger transmission except at the lower energies. 

\begin{figure}[htbp]
\includegraphics[width=12cm]{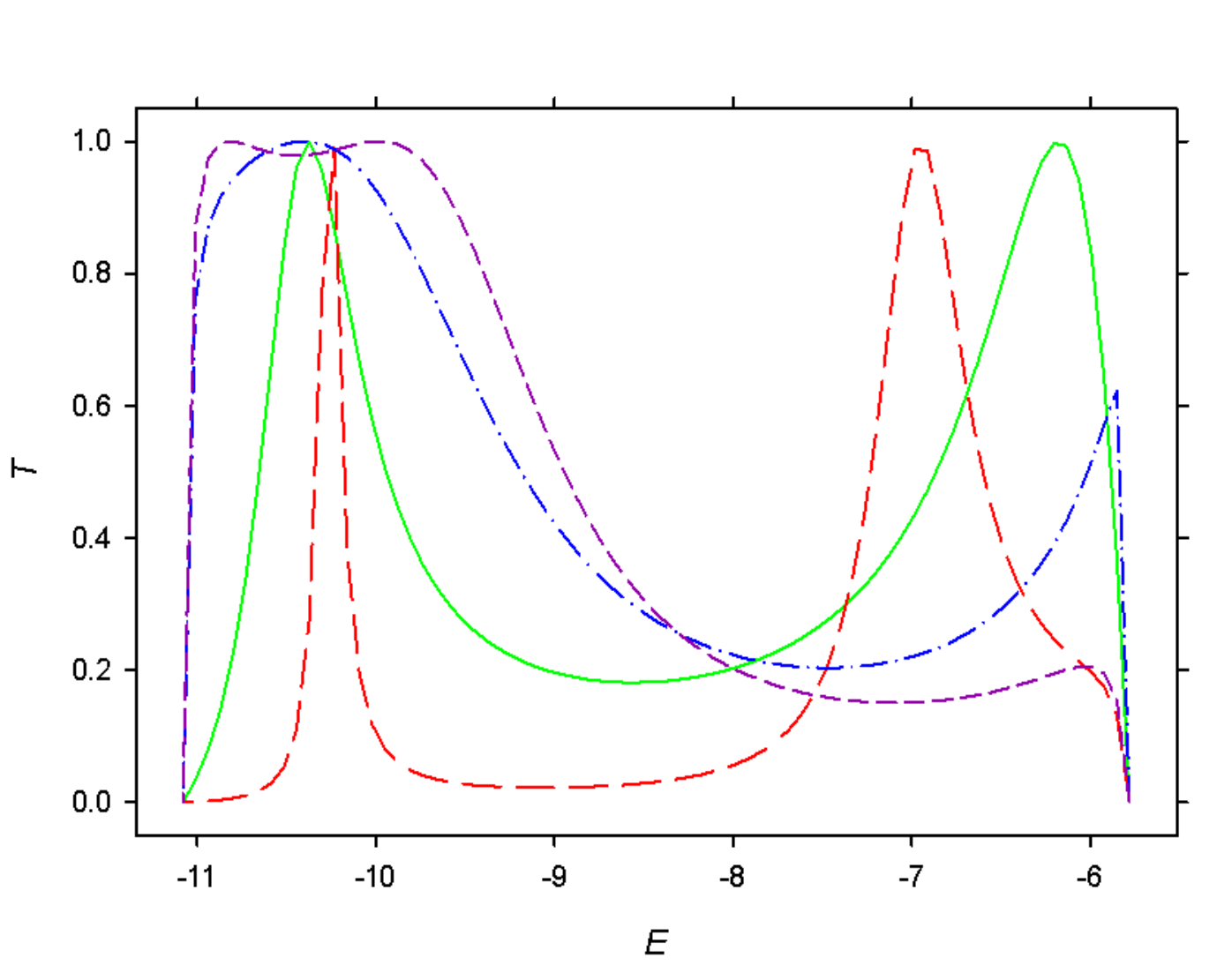}
\caption{Transmission $T$ versus energy $E$ for para-benzene, with $\beta_S=$ (a) $-1.325$ (green solid curve), 
(b) $-0.75$ (red long-dashed), (c) $-1.75$ (blue dash-dotted), (d) $-2.0$ (purple short-dashed). }
\label{fig6}
\end{figure}

Next, we look at the variation of $T(E)$ with $\beta_S$, with reference to Figure \ref{fig6}.
As before, the $T(E)$ curve for the standard parameters, where $\beta_S=-1.325$, is shown in Figure \ref{fig6}(a) 
(same as in Figures \ref{fig4}(a) and \ref{fig5}(a)), as a starting point.
Increasing $\beta_S$ to $-0.75$ (thus weakening the bond), as shown in Figure \ref{fig6}(b), 
produces a somewhat similar curve but with some
noticeable differences, such as a lowering of the local minimum.
The lower peak remains at almost the same energy, while becoming much narrower.
The upper resonance is more drastically changed.
Although it remains resonant, it is shifted to a significantly lower energy, while at its upper-energy side, a noticeable bump has
appeared.
As $\beta_S$ is further increased (not shown), the bump becomes more pronounced and eventually becomes a new resonance
separated from its ``parent'' by a deep local minimum.
As $\beta_S$ tends to 0, this pair of resonances as well as the original lower resonance all tend towards $\delta$-functions,
with $T=0$ at all other energies.
On decreasing $\beta_S$ from its initial value of $-1.325$ to $-1.75$ (see Figure \ref{fig6}(c)), the opposite trend occurs, with
the lower resonance being maintained and indeed broadened, while the upper resonance is destroyed and diminished to a much
smaller height, while shifting to a slightly higher energy.
Further decrement of $\beta_S$ to $-2.0$ (see Figure \ref{fig6}(d)) exacerbates the destruction of the former resonance, with it
now appearing as merely a slightly heightened peak near the upper band edge.
Meanwhile, the lower resonance is beginning to split into a pair of resonances separated by a very shallow minimum.
Further decreases of $\beta_S$ indicate a continuation of these trends (not shown), and in particular, the splitting of the lower
resonance becomes more pronounced.

\subsection{Meta-benzene}

\begin{figure}[htbp]
\includegraphics[width=12cm]{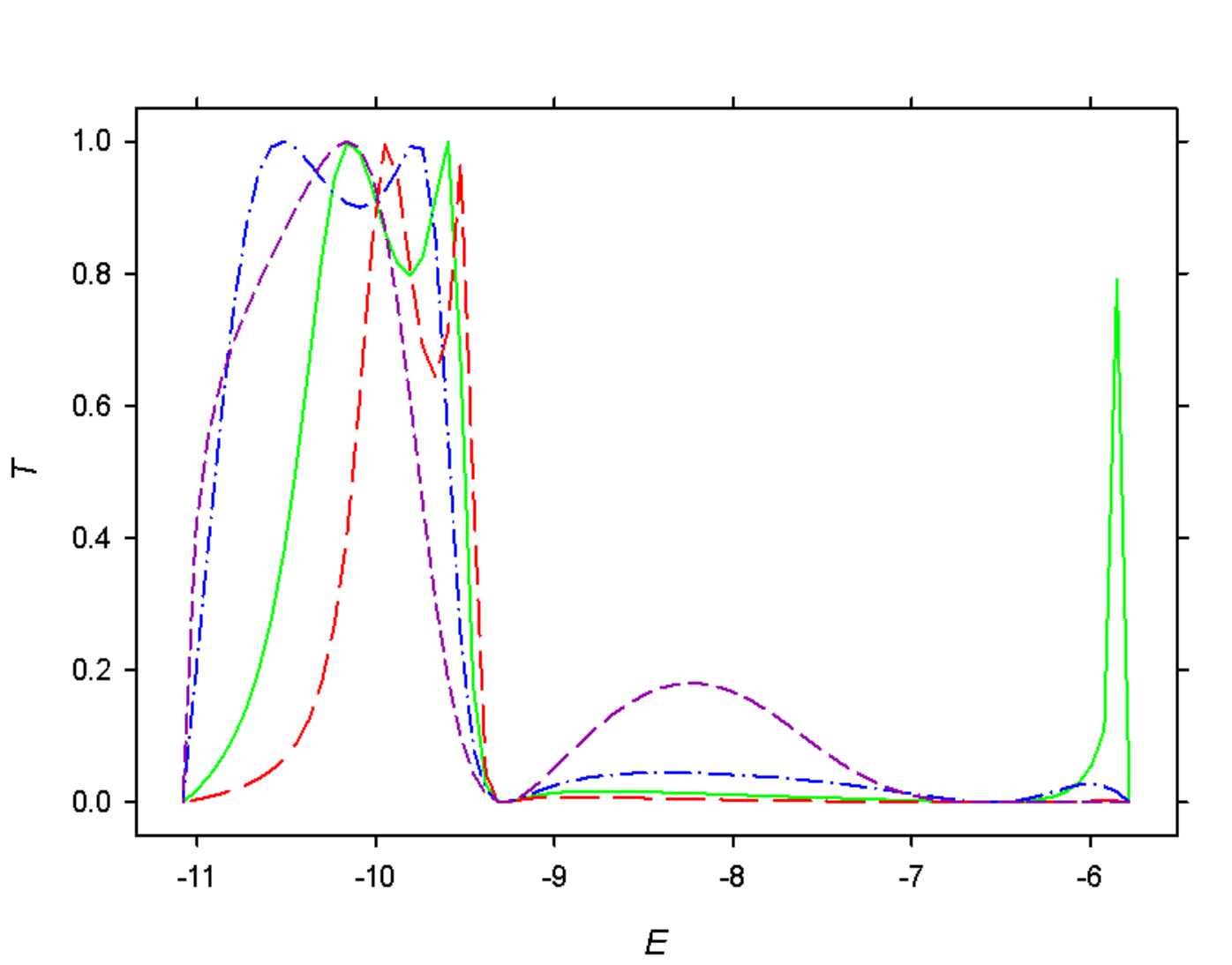}
\caption{Transmission $T$ versus energy $E$ for meta-benzene, with $\alpha_S=$ (a) $-6.553$ (green solid curve), 
(b) $-5.5$ (red long-dashed), (c) $-7.5$ (blue dash-dotted), (d) $-8.5$ (purple short-dashed). }
\label{fig7}
\end{figure}

We now turn to the case of meta-benzene.
As with para-benzene, we use as a reference point the $T(E)$ curve for the standard parameters, which is shown as 
the green solid curve (a) in each of the figures. 
Starting with the dependence on $\alpha_S$, Figure \ref{fig7} shows the variation for the indicated values.
As mentioned, Figure \ref{fig7}(a) displays the $T(E)$ curve for the standard parameters, which include $\alpha_S=-6.553$.
The shape of the curve is generally controlled by a pair of anti-resonances, located at $\alpha=-6.553$ and 
$\alpha+\beta=-9.287$, for which $T=0$.
In general, the meta-benzene molecule allows as many as 4 anti-resonances \cite{ref10}, whose energies are determined solely by the C parameters
$\alpha$ and $\beta$, but their existence depends on their alignment with the energy band of the Au leads, which is 
determined by its parameters $\alpha_{Au}$ and $\beta_{Au}$.
Thus the anti-resonances are pinned in energy with respect to the contact parameters ($\alpha_S$, $\beta_C$
and $\beta_S$) under consideration here. 
(The transmission is also always 0 at the band edges.)
The anti-resonances divide the band in Figure \ref{fig7}(a) into 3 sub-bands, for which the lowest one is clearly the one in
which transmittivity is concentrated.
In that lowest band, transmission is dominated by a pair of resonances, separated by a moderate local minimum.
Transmission is suppressed to near-zero values throughout the middle and upper sub-bands, except for a near-resonance
at the upper band edge.
When $\alpha_S$ is raised (to $-5.5$ as in Figure \ref{fig7}(b)),  transmission is decreased to virtually 0 throughout the middle
and upper sub-bands, including at the upper band edge.
Meanwhile, the lower sub-band looks much the same as in (a), with the resonances shifting somewhat in energy and the local
minimum becoming deeper. 
As $\alpha_S$ is further increased, the local minimum further deepens, and the two resonances tend towards a pair of 
$\delta$-functions, with $T=0$ at all other energies.
When $\alpha_S$ is lowered to $-7.5$ (see Figure \ref{fig7}(c)), the resonances diverge somewhat while their separating
minimum becomes shallower, but still with only modest shifts in energy.
Transmission in the middle sub-band is actually enhanced somewhat, while that in the upper sub-band is again deeply suppressed.
Further lowering of $\alpha_S$ to $-8.5$ (Figure \ref{fig7}(d)) continues these trends, with transmission in the upper sub-band
completely suppressed while that in the middle sub-band is visibly enhanced, to a maximum of $T \approx 0.2$.
In the lower sub-band, the local minimum has disappeared, causing the two resonances to merge into a single one, which
persists and shifts to lower energies with further decreases in $\alpha_S$.

\begin{figure}[htbp]
\includegraphics[width=12cm]{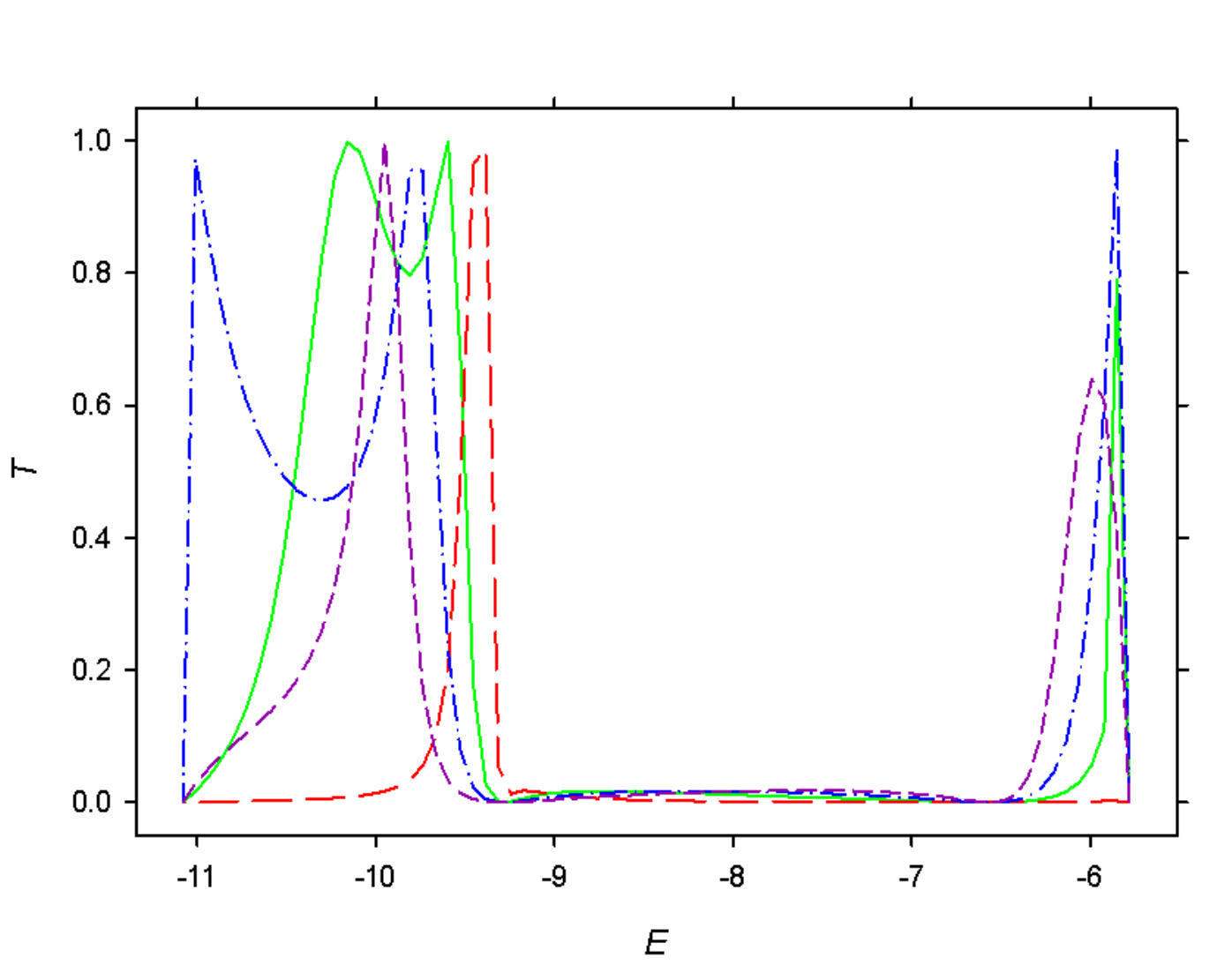}
\caption{Transmission $T$ versus energy $E$ for meta-benzene, with $\beta_C=$ (a) $-2.1872$ (green solid curve), 
(b) $-1.0$ (red long-dashed), (c) $-3.0$ (blue dash-dotted), (d) $-4.0$ (purple short-dashed). }
\label{fig8}
\end{figure}

We turn now to the dependence of meta-benzene on parameter $\beta_C$, shown in Figure \ref{fig8}, with as usual (a) reproducing
the $T(E)$ curve for the standard parameters (including $\beta_C=-2.1872$) from Figure \ref{fig7}(a).
Decreasing $|\beta_C|$, for a weaker bond, causes the 2 resonances in the lower sub-band to move closer together, while
raising their adjoining minimum,  as in Figure \ref{fig8}(b), until they eventually coalesce into a single resonance, which then
tends towards a $\delta$-function as $\beta_C$ moves closer to 0.
Meanwhile, in the other two sub-bands, transmission is reduced to effectively 0.
Increasing $|\beta_C|$ to make a stronger bond, as in Figure \ref{fig8}(c), has little effect on the middle sub-band, with 
$T$ remaining non-zero but very low.
In the upper sub-band, the peak in $T$ near the upper band edge is enhanced and broadened into a resonance.
The lower sub-band remains the region of greatest transmittivity, with the 2 resonances further separating with a deepening
of the local minimum, and the lower resonance moving close to the lower band edge. 
Further strengthening of that bond (to $\beta_C=-4.0$ in Figure \ref{fig8}(d)) has the effect of shifting the resonances to
lower energies, so much so that the one at the lower band edge vanishes by being shifted below the band. 
Meanwhile, the peak near the upper band edge undergoes sufficient broadening that it loses resonance ($T<1$) and is shifted to
a somewhat lower energy, while transmission in the middle sub-band remains negligible.

\begin{figure}[htbp]
\includegraphics[width=12cm]{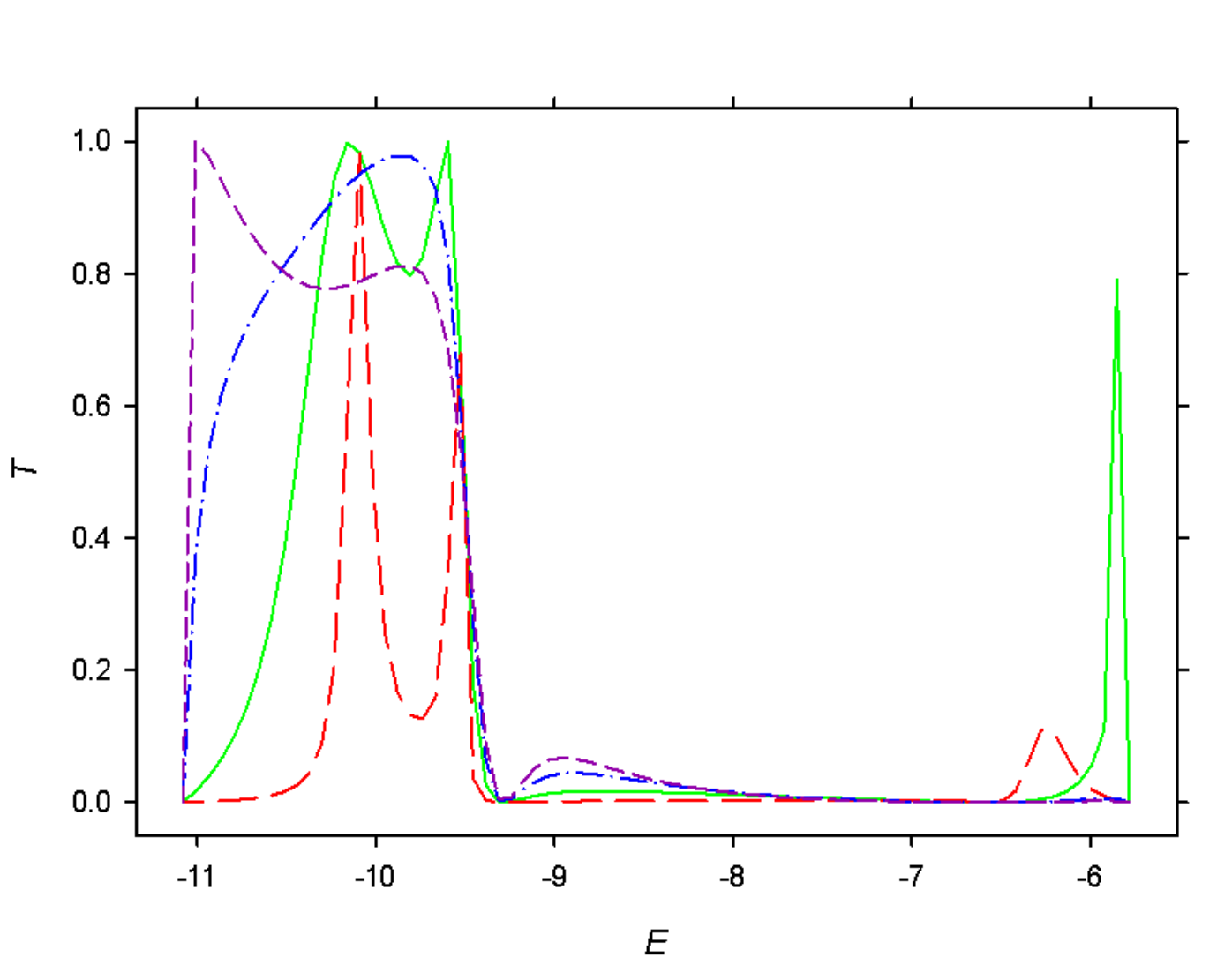}
\caption{Transmission $T$ versus energy $E$ for meta-benzene, with $\beta_S=$ (a) $-1.325$ (green solid curve), 
(b) $-0.75$ (red long-dashed), (c) $-1.75$ (blue dash-dotted), (d) $-2.0$ (purple short-dashed). }
\label{fig9}
\end{figure}

Lastly for this subsection, we look at the dependence of the transmission on $\beta_S$ in Figure \ref{fig9}, with (a) showing
the curve for the standard parameters, including $\beta_S=-1.325$.
Raising $\beta_S$ to $-0.75$, as in Figure \ref{fig9}(b), reduces to 0 the transmittivity in the middle sub-band and greatly
diminishes it in the upper sub-band (maximum $T \approx 0.1$).
The pair of resonances in the lower sub-band persist, but are narrower and separated by a deeper minimum, which as 
$\beta_S$ increases further towards 0, tend towards a pair of $\delta$-functions, the sole regions of transmission in the band.
Decreasing $\beta_S$ to create a stronger bond, however, has the effect in the lower sub-band of coalescing the two resonances into one peak with $T$
slightly less than 1 (Figure \ref{fig9}(c)) while marginally enhancing transmission in the middle sub-band and diminishing it to virtually 0
in the upper one.
Further decreases in $\beta_S$ (Figure \ref{fig9}(d)) continue these trends in the middle and upper sub-bands, while in the lower
one, the peak is shifted to a lower energy, close to the lower band edge with a side peak emerging near the anti-resonance at $E=-9.287$.
This side peak in the lower sub-band appears to be contributing, in a sense, to the enhancement of transmission in the middle sub-band, albeit despite the interference caused by the pinned anti-resonance between them.

\subsection{Ortho-benzene}

\begin{figure}[htbp]
\includegraphics[width=12cm]{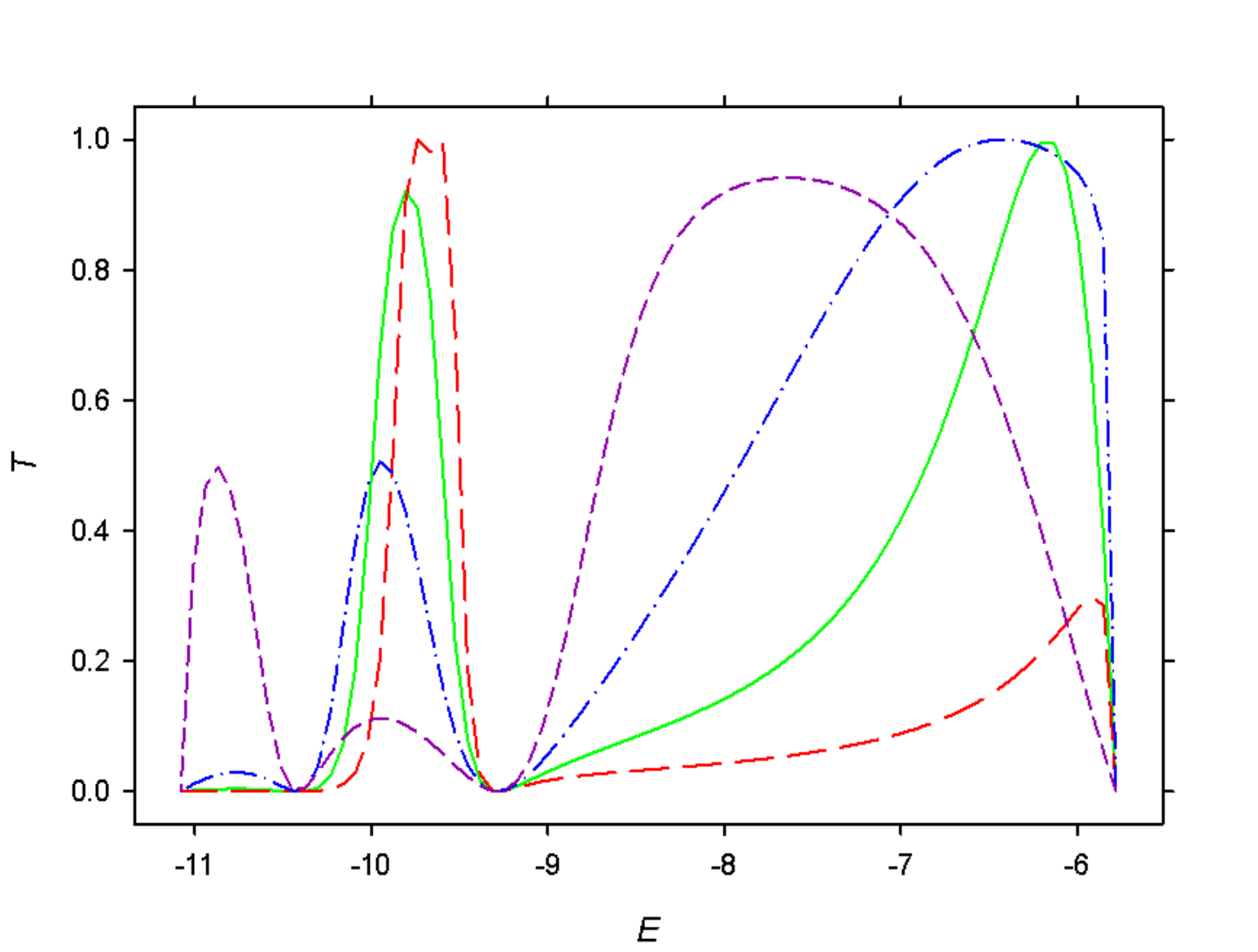}
\caption{Transmission $T$ versus energy $E$ for ortho-benzene, with $\alpha_S=$ (a) $-6.553$ (green solid curve), 
(b) $-5.5$ (red long-dashed), (c) $-7.5$ (blue dash-dotted), (d) $-8.5$ (purple short-dashed). }
\label{fig10}
\end{figure}

Lastly, we consider the case of ortho-benzene. 
Figure \ref{fig10}(a) shows the $T(E)$ curve for the standard parameters, which as usual, is used as a reference point
when varying the parameters.
Similarly to meta-benzene, ortho-benzene in principle admits as many as 4 anti-resonances \cite{ref10}, depending on the specific values of
$\alpha$, $\beta$, $\alpha_{Au}$ and $\beta_{Au}$, but independent of the contact parameters $\alpha_S$, $\beta_C$
and $\beta_S$. 
For the parameters adopted here, there are only 2 anti-resonances, which are pinned at $\alpha+\beta=-9.287$ and
$\alpha+ \sqrt2 \beta=-10.419$, so $T=0$ at these energies, for all cases considered here.
(The transmission is also always 0 at the 2 band edges.)
The anti-resonances thus divide the band into 3 sub-bands as in Figure \ref{fig10}(a).
The lowest sub-band shows near-zero transmission throughout, while the middle sub-band shows a peak of $T \approx 0.92$ at
$E=-9.8$ and the upper one is dominated by a resonance of $T=1$ at $E=-6.2$.
On raising $\alpha_S$ to $-5.5$ (Figure \ref{fig10}(b)), the transmission in the lowest sub-band remains virtually 0.
In the middle sub-band, the peak is actually heightened and split into a pair of resonances, separated by a very slight minimum,
and shifted to slightly higher energies. 
Meanwhile, in the upper sub-band, its resonance is noticeably suppressed with maximum $T \approx 0.3$ and the peak shifted
almost to the upper band edge. 
Lowering $\alpha_S$ to $-7.5$ (Figure \ref{fig10}(c)) produces some very modest increase in $T$ in the lowest sub-band.
The peak in the middle sub-band is substantially diminished, and shifted to a slightly lower energy, so transmission is suppressed.
In the upper sub-band, the resonance is maintained but substantially broadened, at a slightly lower energy, resulting in significant
enhancement of transmission across this region.
These trends are generally continued with further lowering of $\alpha_S$, although transmission in the lowest sub-band becomes
clearly greater than in the middle one, and the peak in the upper sub-band is no longer resonant but still high enough
(maximum $T \approx 0.94$) to ensure strong transmission there.

\begin{figure}[htbp]
\includegraphics[width=12cm]{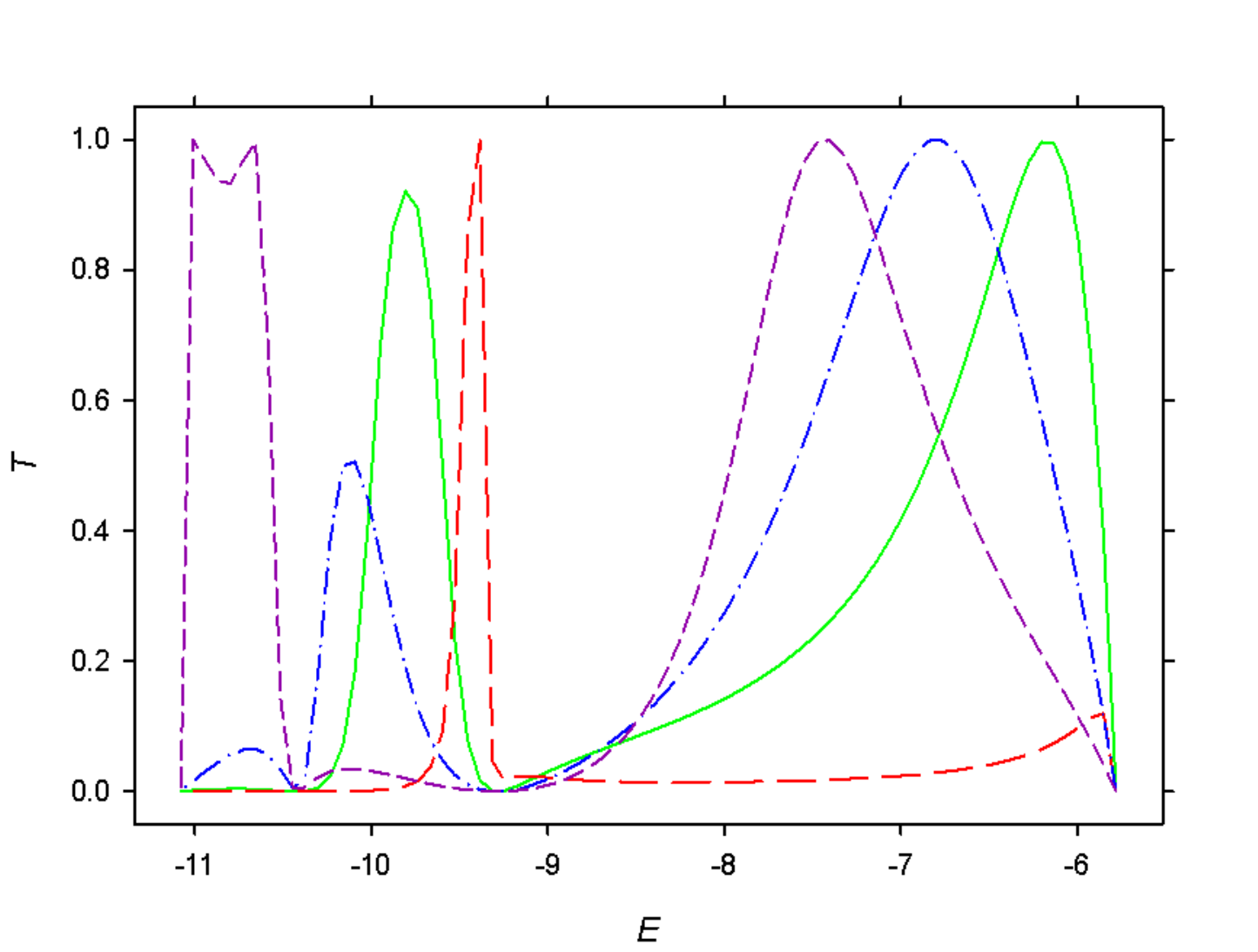}
\caption{Transmission $T$ versus energy $E$ for ortho-benzene, with $\beta_C=$ (a) $-2.1872$ (green solid curve), 
(b) $-1.0$ (red long-dashed), (c) $-3.0$ (blue dash-dotted), (d) $-4.0$ (purple short-dashed). }
\label{fig11}
\end{figure}

Next we turn to the dependence of ortho-benzene on $\beta_C$, exhibited in Figure \ref{fig11}, and with Figure \ref{fig11}(a)
reproducing (from Figure \ref{fig10}(a)) the reference $T(E)$ curve for the standard parameters (for which $\beta_C=-2.1872$).
Weakening the bond by changing $\beta_C$ to $-1.0$, shown in Figure \ref{fig11}(b), retains transmission in the lowest sub-band
at negligible levels, and dramatically reduces it in the upper one so that $T < 0.12$.
Meanwhile in the middle sub-band, the dominant peak is heightened and narrowed into a resonance at $E \approx -9.4$, providing only
this small region of significant transmission in the entire band.
Continued reduction in $|\beta_C|$ towards $0$ (not shown) tends this resonance towards a $\delta$-function, with very low 
transmission in the upper sub-band, and effectively none in the lower one.
On the other hand, strengthening the bond by changing $\beta_C$ to $-3.0$ (Figure \ref{fig11}(c)) noticeably increases 
transmission in the lowest sub-band, but only to a maximum of $T \approx 0.065$.
The peak in the middle sub-band is reduced in height by about half, with a modest shift in energy.
The resonance in the upper sub-band is maintained, but shifted to a lower energy while being broadened, thus enabling generally
stronger transmission in that region.
Further strengthening of the bond (Figure \ref{fig11}(d) where $\beta_C = -4.0$) mostly continues these effects, with the resonance
in the upper sub-band shifting to a still lower energy, and the peak in the middle one diminished to an extremely low level.
The most dramatic change is in the lowest sub-band where transmission is substantially increased, with the $T(E)$ curve now displaying
a pair of resonances, separated by a shallow minimum, so that region enjoys high transmission.

\begin{figure}[htbp]
\includegraphics[width=12cm]{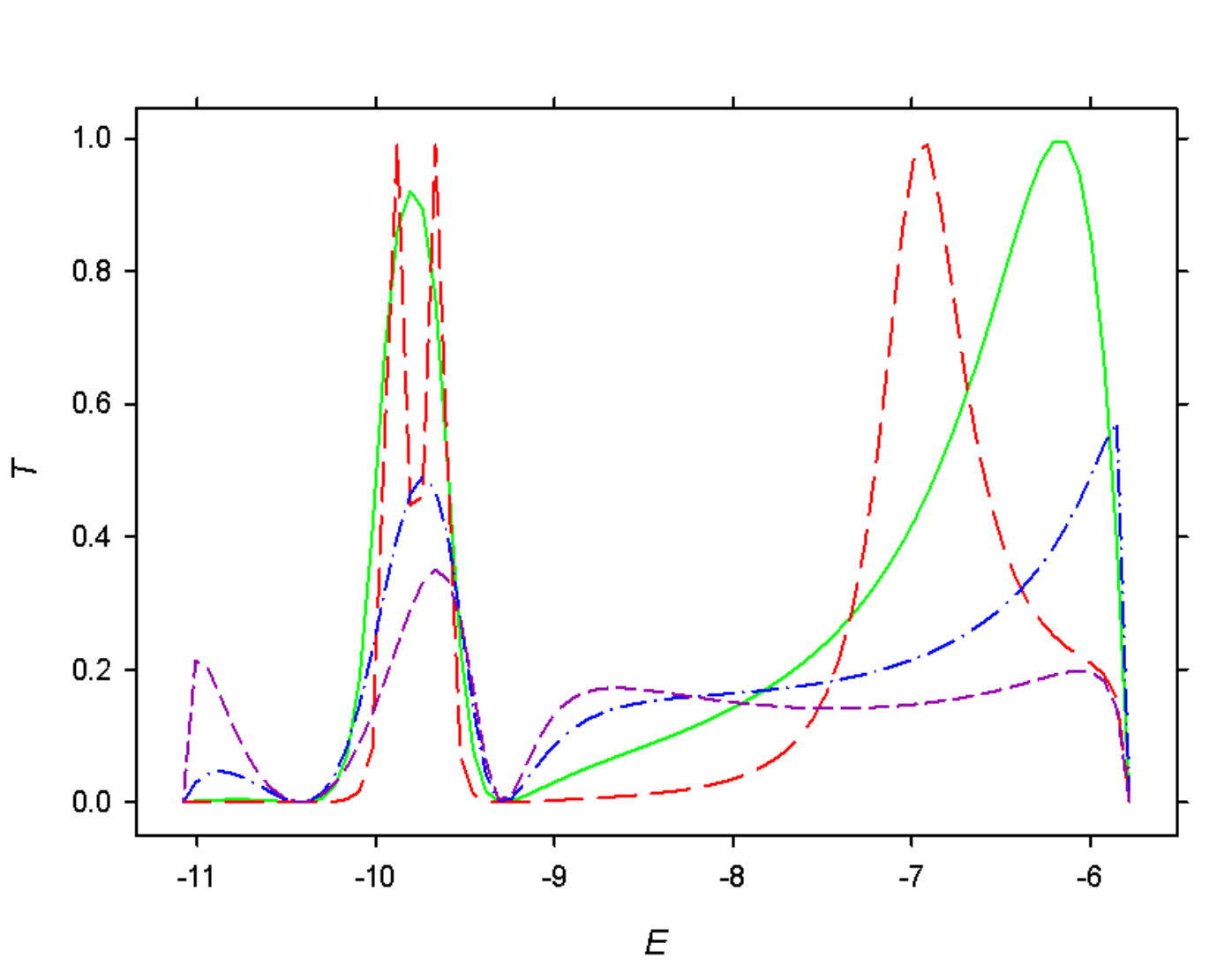}
\caption{Transmission $T$ versus energy $E$ for ortho-benzene, with $\beta_S=$ (a) $-1.325$ (green solid curve), 
(b) $-0.75$ (red long-dashed), (c) $-1.75$ (blue dash-dotted), (d) $-2.0$ (purple short-dashed). }
\label{fig12}
\end{figure}

Lastly, we examine ortho-benzene and its dependence on $\beta_S$, with results shown in Figure \ref{fig12}.
Figure \ref{fig12}(a) reproduces the $T(E)$ curve for the standard parameters, which include $\beta_S=-1.325$, already
seen in Figures \ref{fig10}(a) and \ref{fig11}(a).
When $\beta_S$ is raised to $-0.75$ to weaken the bond (see Figure \ref{fig12}(b)), the transmission in the lowest sub-band
remains effectively 0, while that in the upper one remains dominated by the resonance, now shifted to $E \approx -6.9$, with
just a bit of a ``bump'' visible near the upper band edge.
The peak in the middle sub-band has split into a pair of narrow neighbouring resonances, separated by a deep local minimum.
As $\beta_S$ is raised further towards 0 (not shown), these two resonances tend towards $\delta$-functions, as does the resonance
in the upper sub-band, while the ``bump'' resolves itself into a fourth resonance that also tends to a $\delta$-function.
These 4 $\delta$-functions are the only regions of non-zero transmission in the small-$|\beta_S|$ weak-bond regime.
Conversely, decreasing $\beta_S$, first to $-1.75$ (shown in Figure \ref{fig12}(c)) has very different effects.
The upper sub-band resonance is greatly reduced in height, so that the maximum $T$ is approximately $0.57$, while the
curve increases in the lower part of that sub-band, to produce more uniform if modest transmission in that region.
Meanwhile, the peak in the central sub-band persists but is greatly reduced in height, while some noticeable transmission 
appears in the lower sub-band.
All of these trends continue with further decreases in $\beta_S$ ($=-2.0$ in Figure \ref{fig12}(d)).
The transmission improves to a modest maximum of approximately $0.21$ in the lowest sub-band, it is further reduced
in the middle one, and becomes even more uniform in the upper one.

\section{Conclusions}

In summary, we have looked at how transmission through a benzene ring depends on the parameters of the contact
atoms between the ring and the leads.
Specifically, these are the site energy $\alpha_S$ of the contact atom, and its bond energies $\beta_C$ and $\beta_S$
with the nearest atoms of the ring and the leads, respectively.
For all three of the benzene configurations, these parameters are seen to play important roles in determining the 
transmission properties of the system.
In comparing the three configurations, it generally appears that para-benzene is the most sensitive to the parameter values,
due to the lack of anti-resonances within the band, which allows for greater variation of the $T(E)$ curve as a parameter is changed.
Although the meta and ortho configurations each admit two anti-resonances, pinned in energy,
their more equal spacing in meta-benzene would seem to make it the less sensitive of the two to parameter values.
(Compare the degree of variation in Figure \ref{fig4} versus Figure \ref{fig7} versus Figure \ref{fig10}.)

In considering the relative importance of the three parameters, it is difficult to draw definitive conclusions, as all three 
play significant roles. 
Nonetheless, based on our figures, it is suggestive that $\beta_S$ has less dramatic effects than the other two parameters.
This might be expected in light of the fact that $\beta_S$ is the bond energy between the contact atom and the chain, but
not directly connecting to the benzene ring itself.

\section{Keywords}

benzene, electron transmission, parameters, renormalization method

\end{document}